\documentclass[aps,prl,twocolumn,groupedaddress]{revtex4}
\usepackage{graphicx}
\usepackage{epsfig}
\begin{document}

\title{Strain Hardening in Polymer Glasses: Limitations of Network Models}
\author{Robert S. Hoy}
\email{robhoy@pha.jhu.edu}
\author{Mark O. Robbins}
\affiliation{Department of Physics and Astronomy, Johns Hopkins University, Baltimore, MD 21218}

\date{May 29, 2007}
\begin{abstract}
Simulations are used to examine the microscopic
origins of strain hardening in polymer glasses.
While traditional entropic network models can be fit to the total stress,
their underlying assumptions are inconsistent with
simulation results.
There is a substantial energetic contribution to the stress that
rises rapidly as segments between entanglements are pulled taut.
The thermal component of stress is less sensitive to entanglements,
mostly irreversible, and
directly related to the rate of local plastic arrangements.
Entangled and unentangled chains show the same strain hardening
when plotted against the microscopic chain orientation rather than the
macroscopic strain.
\end{abstract}
\maketitle

The stress needed to deform a polymer glass increases as the strain rises.
This strain hardening plays a critical role in stabilizing polymers against
strain localization and fracture,
and reduces wear \cite{mergler04}.
While models have had some success in fitting experimental data,
fundamental inconsistencies in fit parameters and trends imply
that our understanding of the microscopic origins of strain hardening
is far from complete.

Most theories of strain hardening \cite{haward68,arruda93b}
are based on rubber elasticity theory \cite{treloar75}.
These entropic network models
assume that polymer glasses behave like crosslinked rubber,
with the number of monomers between crosslinks equal to the entanglement
length $N_e$.
The increase in the stress $\sigma$ due to deformation by a stretch
tensor $\bar \lambda$ is attributed to the decrease in entropy as
polymers are stretched between affinely displaced entanglements.
One finds \cite{arruda93b}
\begin{equation}
\sigma(\bar\lambda) = \sigma_{0} + G_{R}g(\bar\lambda)L^{-1}(h)/3h
\label{eq:langevinmodel}
\end{equation}
where $\sigma_{0}$ is the yield or plastic flow stress, $G_{R}$ is the strain
hardening modulus, $L^{-1}$ is the inverse Langevin function,
$g(\bar\lambda)$ describes the entropy reduction for
Gaussian chains, and $L^{-1}(h)/3h$ corrects for the finite length
of segments between entanglements.
The value of $N_{e}$ enters in $h$, which is the ratio
of the Euclidean distance between entanglements to the contour length.

Stress-stretch curves for a wide variety of glassy polymers
can be fit to Eq. \ref{eq:langevinmodel},
but the fit parameters are not consistent with the microscopic
picture underlying entropic network models \cite{kramer05}.
For example, values of $N_e$ from fitting $h$ may be several
times smaller than those
obtained from the plateau modulus $G_{N}^0$ \cite{arruda93b}.
Entropic network models predict $G_R \approx G_{N}^0$
near $T_g$, but measured $G_R$ are about 100 times larger \cite{vanMelick03}.
$G_R$ also rises as $T$ decreases \cite{vanMelick03,hoy06},
while any entropic stress must drop to zero as $T \rightarrow 0$
\cite{kramer05}.
Recent work \cite{dupaix05,hoy06} shows that changes in $G_R$
correlate with those in the plastic flow stress. 
Indeed entire stress-stretch curves
collapse when normalized by $\sigma_{0}$ \cite{hoy06}.
This is not expected from entropic models, where $\sigma_0$ is treated
as an independent parameter arising from local plasticity.
A more conceptual difficulty in entropic models is that,
unlike rubber, glasses are not able to dynamically sample chain conformations.

In this Letter we use simulations to examine the microscopic origins
of strain hardening.
While our results for the total stress can be fit to Eq. \ref{eq:langevinmodel},
network models are not consistent with observed
changes in energy, heat flow and molecular conformations.
The stress can be divided into energetic and thermal contributions.
The energetic contribution is strictly zero in the entropic model, but we
find it becomes significant as the segments between entanglements
are stretched taut.
In contrast, entanglements have little direct influence on the thermal
contribution.
This thermal stress is found to be directly related to the rate of
local plastic rearrangements.
Finally,
network models only predict strain hardening for entangled chains ($N \gg N_e$),
yet
substantial hardening is observed for $N$ as small as $N_e/4$.
Results for entangled and unentangled chains collapse when
plotted as a function of the microscopic strain-induced orientation of chains
rather than the macroscopic strain.

Much of the physics of polymer glasses is independent of chemical detail
\cite{haward97,arruda93b,haward93}.
We thus employ a simple coarse-grained bead-spring model \cite{kremer90}
that captures the key physics of linear homopolymers.
Each polymer chain contains $N$ beads of mass $m$.
All beads interact via the truncated and shifted Lennard-Jones potential
$U_{LJ}(r) = 4u_{0}[(a/r)^{12} - (a/r)^{6} - (a/r_{c})^{12} + (a/r_c)^{6}]$,
where $r_{c}=1.5a$ is the cutoff radius and $U_{LJ}(r) = 0$ for $r > r_{c}$.
We express all quantities in terms of the molecular diameter $a$, binding
energy $u_{0}$, and characteristic time $\tau_{LJ} = \sqrt{ma^{2}/u_{0}}$. 

Covalent bonds between adjacent monomers on a chain are modeled using the
finitely extensible nonlinear elastic potential
$U(r) = -(1/2)(kR_{0}^2) ln(1 - (r/R_{0})^{2})$, with the canonical
parameter choices $R_{0} = 1.5a$ and $k = 30u_{0}/a^{2}$\ \cite{kremer90}.   
Chain stiffness is introduced
through a bending potential $U_{bend}(\theta) = k_{bend}(1 - cos\theta)$,
where $\theta$ is the angle between consecutive covalent bond vectors along a
chain.  
Stiffer chains have lower entanglement lengths.
Values of $N_e$ obtained from primitive path analysis (PPA) \cite{everaers04}
range from $N_{e} = 71$ for flexible chains ($k_{bend}=0$) to $N_{e}=22$
for $k_{bend}=2.0u_{0}$. 

Glassy states were obtained by rapid temperature quenches
from well-equilibrated melts \cite{auhl03}
to a temperature $T$ below the glass transition
temperature $T_g \simeq 0.35 u_0/k_B$ \cite{rottler03c}.
While quench rate affects the initial yield stress \cite{rottler05},
it had little influence on strain hardening.
Periodic boundary conditions were imposed, with periods $L_i$ along
directions $i=x$, $y$, and $z$.
The initial periods $L_i^{0}$ were chosen to give zero pressure at $T$.
A Langevin thermostat with damping rate $1/\tau_{LJ}$
was applied
to the peculiar velocities 
in all three directions.

Experiments commonly impose compressive deformations because
they suppress strain localization \cite{vanMelick03,haward93,hasan93}.
Simulations were performed for both uniaxial and plane strain compression.
Both show the same behavior, and only uniaxial results are presented below.
The cell is compressed along $z$ 
at constant true strain rate $\dot{\epsilon} = \dot{L_{z}}/L_{z}$.
Results for $\dot\epsilon=-10^{-5}/\tau$ are shown below, but
similar behavior is found at $\dot\epsilon=-10^{-3}/\tau$.
Qualitative changes can occur at the higher rates employed in recent
atomistic simulations of strain hardening \cite{lyulin04,lyulin05,li06,foot1}.
The stress perpendicular to the compressive axis
is maintained at zero by varying $L_x$ and $L_y$ \cite{yang97}.
Fits to network models normally assume that the volume remains constant
and $L_x=L_y$
and this is approximately true in our simulations.
Then deformation can be expressed in terms of a single stretch component
$\lambda \equiv L_z/L_z^0$.
  
\begin{figure}[h!]
\includegraphics[width=3.2in]{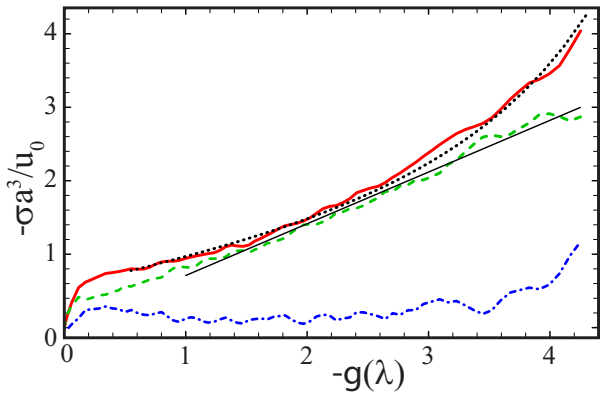}
\caption{(Color online) Total stress (solid line) and contributions
from heat (dashed line) and potential energy (dot-dashed line)
for a system with $T=0.2u_0/k_B$, $N=350$ and $N_e=22$.
Fits of $\sigma$ to Eq. \ref{eq:langevinmodel} with $N_e=13$ (dotted line)
and of $\sigma_Q$ to a straight line are also shown.
Both $\sigma$ and $g$ are negative under compression.}
\label{fig:uniaxialthreestresses}
\end{figure}

Typical strain hardening results are shown in Fig.
\ref{fig:uniaxialthreestresses}.
As in experiments,
the stress is plotted against $g(\lambda) \equiv \lambda^2 - 1/\lambda$.
Then entropic network models (Eq. \ref{eq:langevinmodel})
attribute curvature in the strain hardening regime
to reductions in entropy associated with the finite length $N_e$
between entanglements.
The strong upward curvature in Fig. \ref{fig:uniaxialthreestresses}
can be fit to Eq. \ref{eq:langevinmodel} (dotted line), but
with a value of $N_e=13$ that is much smaller than that determined
from $G^0_N$ or PPA ($N_e=22$).
Similar reductions are found for other chain stiffnesses and
in fits to experiment.

The stress represents the incremental work done on a unit volume
of the system by an incremental strain.
It can be divided into contributions from changes in the
internal energy density $U$ and the heat flow out of a unit volume $Q$:
$\sigma= \sigma_U+\sigma_Q$ where $\sigma_U= \partial U/\partial \epsilon$
and $\sigma_Q=\partial Q/\partial \epsilon$.
The derivation of Eq. \ref{eq:langevinmodel} assumes that
$\sigma_U$ does not contribute to strain hardening and that
$\sigma_Q$ is entirely due to reversible
heat associated with changes in entropy.
Simulations allow these assumptions to be tested.

Figure \ref{fig:uniaxialthreestresses} shows that results for $\sigma_Q$
can be fit to the linear behavior predicted for the entropy of ideal
Gaussian chains at $|g|>1$.
Fits to smaller $|g|$ can be obtained with $N_e = 30$
in Eq. \ref{eq:langevinmodel}, but fits
to uniaxial and plane strain results always give $N_e$ that are
larger than values obtained from $G_N^0$ and PPA,
and much larger than values from fits to the total stress.
Separate simulations show that $\sigma_Q$ is dominated by irreversible
heat flow rather than changes in entropy.
After straining to a large $|g|$ the stress is returned to zero.
The stretch only relaxes about 10\% and only $\sim 5$\% of the work
associated with $\sigma_Q$ is recovered.
Similar irreversibility is observed in experiments \cite{hasan93},
confirming that the force can not be entropic.

The energetic contribution to the stress in Fig. \ref{fig:uniaxialthreestresses}
is important during
the initial rise to the plastic flow stress $\sigma_0$.
The value of $\sigma_U$ then
drops to a small constant for $0.5 < |g| < 2.5$.
At larger strains there is a pronounced upturn in $\sigma_U$
that contributes almost all of the curvature in the total stress.
This energetic term thus has a crucial effect on fit values of $N_e$
even though the derivation of Eq. \ref{eq:langevinmodel} assumes $\sigma_U=0$.
Similar results are found for all $T$ and $k_{bend}$,
and for uniaxial and plane strain.
In all cases, $\sigma_Q$ exhibits nearly ideal Gaussian behavior
($L^{-1}(h)/3h \simeq 1$) and $\sigma_U$ leads to a more rapid rise
in stress at large stretches.
The effect of $\sigma_U$ increases and extends to smaller $|g|$
as the intrinsic $N_e$ from PPA decreases.

Examination of the evolving conformations of individual
chains also provides tests of entropic network
models.
If entanglements act like crosslinks, then polymer glasses
should deform affinely on scales greater than the entanglement spacing.
Our recent studies confirm this affine displacement, and the
associated increase in $h$ as segments between entanglements pull taut
\cite{hoy06}.
Fig. \ref{fig:uniaxialthreestresses} shows that this increase in $h$
has little effect on the thermal terms that motivated
Eq. \ref{eq:langevinmodel}.
Instead, straightening of segments produces large energetic terms
by disrupting the local packing structure.
Energy is stored in increasing tension in the covalent bonds
countered by compression of intermolecular bonds.
Experiments also find significant energy storage \cite{hasan93,chui99},
and could in principal track $\sigma_U$ over the full strain
hardening regime.

\begin{figure}[htbp]
\includegraphics[width=3.2in]{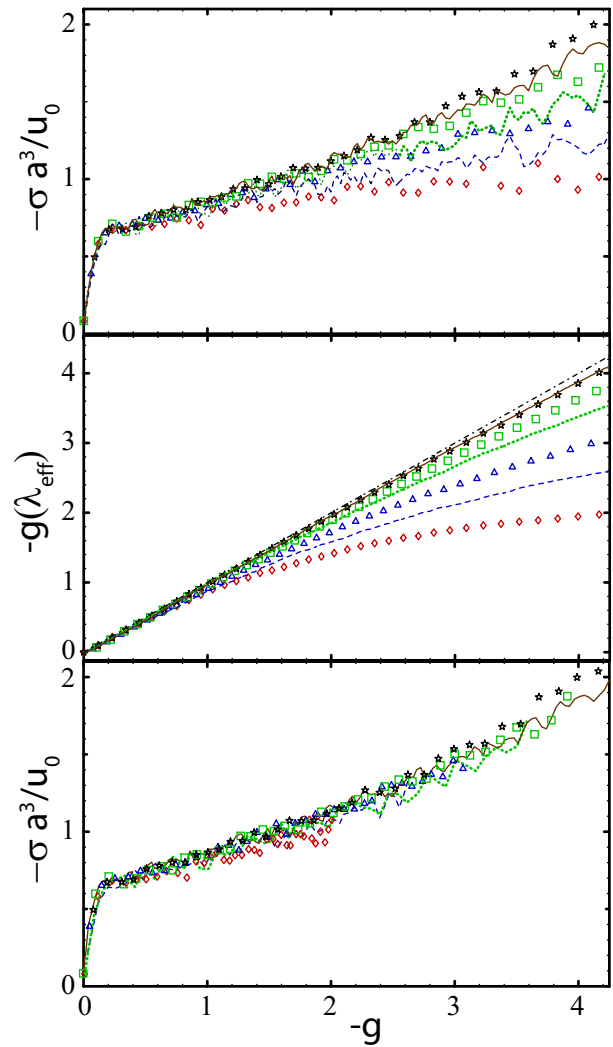}
\caption{
(Color online)
(a) Stress vs. $g(\lambda)$ during uniaxial compression at $k_B T=.2 u_{0}$
for $k_{bend}=0.75 u_0$, $N_e = 39$ and
strain rate $\dot\epsilon = -10^{-5}/\tau_{LJ}$.
Successive curves from bottom to top are for
N = 10 ($\diamond$), 16 ($- - -$), 25 ($\triangle$),
40 ($\cdots$),
70 (squares),
175 ($\Huge -$), and
350 ($\star$).
(b) $g(\lambda_{eff})$ vs. $g(\lambda)$ for the same systems.
The dot-dashed line corresponds to $\lambda_{eff} = \lambda$.
(c) Stresses for different $N$ collapse when plotted against $g(\lambda_{eff})$.
}
\label{fig:collapse}
\end{figure}

Further insight into strain hardening is provided by examining
the dependence on chain length.
Entropic network models assume that the length
should not matter for highly entangled systems, 
$N\gg N_e$, and there should be no network
to produce strain hardening for $N< N_e$.
Simulations confirm that $\sigma$ is independent of $N$ for
$N \gg N_e$, but show substantial strain hardening for
$N < N_e$ \cite{hoy06,lyulin04}.
Figure \ref{fig:collapse}(a) illustrates this hardening for
chains as short as $N_e/4$.
Results for short chains follow
the asymptotic behavior of highly entangled chains
($N/N_e > 4$) to larger $|g|$ as $N$ increases.
This suggests that deformation changes chain conformations on longer
scales as $|g|$ increases and that entanglements only become relevant
at large $|g|$ ($|g| \gtrsim 2.5$ in Fig. \ref{fig:collapse}).

The observation of strain hardening implies that the microscopic
arrangement of chains evolves under stretching.
One way of quantifying this is through changes in the root mean squared
components $R_{i}$ of the end-to-end vectors of chains relative
to their initial values $R_{i}^0$.
Under an affine deformation at constant volume:
$\lambda=R_{z}/R_{z}^0 =(R_{x}^0/R_x)^2=(R_{y}^0/R_{y})^2$.
The deformation of short chains is subaffine\cite{dettenmaier86,hoy06},
but we find that the above ratios \cite{foot3b} 
can all be described by an effective stretch $\lambda_{eff}$
\cite{foot3}.
Figure \ref{fig:collapse}(b) shows $g(\lambda_{eff})$ as
a function of $g(\lambda)$ for different $N$.
All chains show significant stretching, and
highly entangled systems deform affinely.
The small deviation between g($\lambda$) and g($\lambda_{eff}$) for $N\gg N_e$
results from a small increase in density ($\sim 4$\%)
at large $|g|$ rather than nonaffine deformation.
As $N$ decreases, the deformation becomes subaffine at smaller and
smaller $|g|$.
This confirms that the scale over which chain conformations are
distorted grows with $|g|$, and that entanglements only
become important for $|g| > \gtrsim 2.5$ in this system.

Strain hardening is directly related to the changes in chain
conformation represented by $\lambda_{eff}$ rather than the
macroscopic deformation $\lambda$ \cite{foot3}.
To illustrate this, data from Fig. \ref{fig:collapse}(b)
are replotted against $g(\lambda_{eff})$ in Fig. \ref{fig:collapse}(c).
Data for different chain lengths collapse onto a single curve
even though $N$ is as much as 4 times smaller than $N_e$.
Similar results are found for other $T$, $N_e$ and for plane strain
compression.
Deviations are only seen when $N$ becomes comparable to the
persistence length and chains can no longer be viewed as
Gaussian random walks.
These results show that entanglements do not have a direct effect
on strain hardening.
Their main role appears to be in forcing the local stretching of
chains $\lambda_{eff}$ to follow the global stretch $\lambda$.

The recently observed \cite{hoy06,dupaix05} correlation between
the strain hardening modulus $G_R$ and
the plastic flow stress $\sigma_0$ suggests that local plastic
rearrangements dissipate most of the energy during compression.
To monitor the rate of plasticity $P \equiv \delta f/ \delta \epsilon$,
we counted the fraction $\delta f$ of Lennard-Jones bonds
with $r<r_c$ whose length changed by more
than 20\% over small intervals in strain $\delta\epsilon =0.005$.
Tests on this and related amorphous models \cite{luan07} show
that this criterion is large enough to exclude elastic deformations,
and that $\delta\epsilon$
is small enough that a given bond does not undergo multiple events.
To eliminate plastic rearrangements associated with equilibrium
aging, the rate of plasticity during deformation was monitored at $T=0$.

\begin{figure}[htbp]
\includegraphics[width=3.2in]{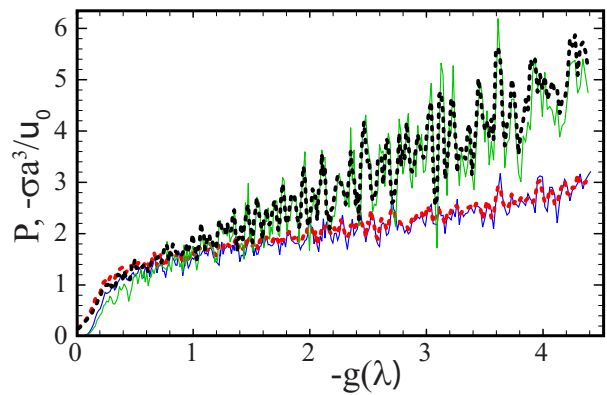}
\caption{(Color online) Rate of plastic rearrangements $P$ (solid lines)
as a function of $g(\lambda)$ for $N_e=26$ (upper curve)
and 71 (lower curve) at $k_B T=0 u_0$.
Dashed lines show the corresponding dissipative stress $\sigma_Q$.
}
\label{fig:bonddamage}
\end{figure}

Figure \ref{fig:bonddamage} shows the rate of plasticity
for $N_e=26$ and $71$.
There is a rapid initial rise as $\sigma$ approaches $\sigma_0$,
followed by a nearly linear increase during the strain hardening regime.
Also plotted in Fig. \ref{fig:bonddamage} are results for $\sigma_Q$.
A fixed vertical rescaling
(coincidentally close to unity in our units)
produces an excellent collapse
of $P$ and $\sigma_Q$ for all $N_e$.
Note that even the 
fluctuations in the quantities are correlated 
\cite{foot4}.
Similar results are found for other criteria for
the rate of plasticity, with only the scaling factor changing.

The above results clearly illustrate that the thermal contribution to
strain hardening is associated
with an increase in the rate of plastic rearrangements as chains stretch.
It remains unclear why this increase should approximately
follow the increase in entropic stress predicted for Gaussian chains.
The entropic stress represents the rate of decrease in the logarithm
of the number of available chain conformations.
One possibility is that the rate of plastic rearrangements scales
in the same way because a decrease in the number of conformations
necessitates larger scale plastic rearrangements.
The relationship between the plastic flow stress and hardening
modulus follows naturally from this picture, and it also
explains why data for different chain lengths
collapse when plotted against $\lambda_{eff}$.
Analytic investigations of this scenario may prove fruitful.

Our results for $\sigma_U$ suggest that the success of Eq.
\ref{eq:langevinmodel} in fitting the total stress may be coincidental,
and explain why fit values of $N_e$ are generally smaller than
intrinsic values from $G_N^0$ and PPA.
It would be interesting to compare trends in the calculated
$\sigma_U$ and $\sigma_Q$ to experimental results.
These could be obtained by differentiating deformation calorimetry
results for the work and heat,
but existing studie s have only extended to 
the plastic flow regime \cite{adams88, salamatina94, oleinik06}.

This material is based upon work
supported by the National Science Foundation under Grant No.\ DMR-0454947.
We thank G. S. Grest for providing equilibrated states and
K. S. Schweizer and S. S. Sternstein for useful discussions.
Simulations were performed with the LAMMPS software package
(http://lammps.sandia.gov).


\begin{thebibliography}{26}
\bibitem{foot1}
At sufficiently high rates, chains are constrained to deform affinely.
This gives $\lambda _{eff}=\lambda $ and the shear stress is nearly
independent of $\mathaccent "705F\relax \epsilon $. Thus the conclusion
that strain hardening is controlled by $\lambda _{eff}$ also applies in
this limit.
\bibitem{foot3b}
All three ratios give the same value of $\lambda _{eff}$ when the chains are long compared to their persistence length. For shorter chains the best correlation is found by using results for the stretched directions.
\bibitem{foot4}
A similar collapse is not achieved for the total stress, which has smaller fluctuations.
\end{thebibliography}
\end{document}